# Extraction of the Eliashberg function from tunneling data. The method and the program

A. I. D'yachenko.

*A. A. Galkin Donetsk Physical and Technical Institute of the National Academy of Sciences of Ukraine.*
*72 R. Luxemburg St., Donetsk 83114, Ukraine*
E-mail: dyachen@hsts.fti.ac.donetsk.ua

Electron-phonon interaction (EPI) is the basis for many phenomena in condensed matter physics [1,2]. In principle, photoemission [3], X-Ray scattering [4] and optical measurements [5] can give a certain information on the EPI, although the main experimental tool for the determination of the spectral function $\alpha^2 F(\omega)$ in superconductors so far is the classical (Giaever) tunneling measurement for which the energy resolution is better than 0.1 meV [6]. The spectral function $\alpha^2 F(\omega)$ can be obtained from the first derivative of the tunnel current $dI/dV$ which for SIN junction at T=0 is simply proportional to the so-called tunnel density of states $N(\omega)$,

$$\left(\frac{dI}{dV}\right)_S \Big/ \left(\frac{dI}{dV}\right)_N = N(\omega) = \mathrm{Re}\left\{\frac{\omega}{\sqrt{\omega^2 - \Delta^2(\omega)}}\right\}\Bigg|_{\omega=eV} \quad (1)$$

where V is the applied voltage, $\Delta(\omega)$ is the complex superconducting gap which depends on energy $\omega$. According to the Eliashberg theory the connection between the EPI function $\alpha^2 F(\omega)$ and the energy gap parameter $\Delta(\omega)$ of a superconductor is described by the system of equations [1,6]

$$\Delta(\omega) = \frac{1}{Z(\omega)} \int_{\Delta_0}^{\omega_c} d\nu \, \mathrm{Re}\left\{\frac{\Delta(\nu)}{[\nu^2 - \Delta^2(\nu)]^{1/2}}\right\}\left[K^+(\omega,\nu) - \mu^*\right],$$

$$[1 - Z(\omega)]\omega = \int_{\Delta_0}^{\infty} d\nu \, \mathrm{Re}\left\{\frac{\nu}{[\nu^2 - \Delta^2(\nu)]^{1/2}}\right\} K^-(\omega,\nu) , \quad (2)$$

where

$$K^\pm(\omega,\nu) = \int_0^{\omega_o} d\omega_q \alpha^2 F(\omega_q)\left(\frac{1}{\omega_q + \nu + \omega + i\delta} \pm \frac{1}{\omega_q + \nu - \omega - i\delta}\right),$$

$Z(\omega)$ is the renormalization function, $\Delta_0$ is the BCS energy gap in the excitation spectrum of the superconductor, $\Delta(\Delta_0)=\Delta_0$, $\mu^*$ and $\omega_c$ are respectively the Coulomb pseudopotential and the cutoff parameter. The effective electron-phonon interaction function $\alpha^2 F(\omega)$ is averaged over the Fermi surfaces and $\omega_0$ is the maximum phonon frequency.

The function $\alpha^2 F(\omega)$ can be viewed as a product of the phonon density of states $F(\omega)$ and an effective electron-phonon matrix-element square $\alpha^2(\omega)=\alpha^2 F(\omega)/F(\omega)$, which represents the strength of the electron-phonon coupling. The dimensionless electron-phonon coupling constant $\lambda$ is defined by

$$\lambda = 2\int_0^{\omega_0} d\omega \frac{\alpha^2 F(\omega)}{\omega}$$

The basis of the method of reconstruction $\alpha^2 F(\omega)$ and $\mu^*$ from the tunnel density of states (1) is the Kramers-Kronig dispersion relation [7]

$$\mathrm{Im}\left\{\frac{\omega}{[\omega^2-\Delta^2(\omega)]^{1/2}}\right\}=\frac{2\omega}{\pi}\int_{\Delta_0}^{\infty}[N(\nu)-N_{BCS}(\nu)]\frac{d\nu}{\omega^2-\nu^2} \quad (3)$$

where

$$N_{BCS}(\omega)=\mathrm{Re}\left\{\frac{\omega}{[\omega^2-\Delta_0^2]^{1/2}}\right\}$$

is the BCS density of states. The relations (3) permits to calculate the function

$$S(\omega)\equiv\frac{\omega}{[\omega^2-\Delta^2(\omega)]^{1/2}}=N_T(\omega)+i\,\mathrm{Im}\left\{\frac{\omega}{[\omega^2-\Delta^2(\omega)]^{1/2}}\right\}$$

and therefore the complex superconducting gap

$$\Delta(\omega)=\pm\omega\left[1-S^{-2}(\omega)\right]^{1/2}. \quad (4)$$

Since now the $\Delta(\omega)$ function is known, we obtain from eq. (2) and $\mathrm{Im}\{\Delta(\omega)Z(\omega)\}$ the linear integral equation for $\alpha^2F(\omega)$:

$$\int_0^{\omega-\Delta_0}\alpha^2F(\omega)P(\omega-\nu)d\nu=\frac{\mathrm{Re}\Delta(\omega)}{\omega}\int_0^{\omega-\Delta_0}\alpha^2F(\nu)N(\omega-\nu)d\nu+$$
$$+\frac{\mathrm{Im}\Delta(\omega)}{\pi}\left[1+\int_0^{\infty}d\nu N(\nu)\int_0^{\omega_0}d\omega_q\frac{2\alpha^2F(\omega_q)}{(\nu+\omega_q)^2-\omega^2}\right] \quad (5)$$

or in the operator form

$$\mathbf{P}\{g\}=\mathbf{B}+\mathbf{C}\{g\}. \quad (5')$$

Here $g\equiv\alpha^2F(\omega)$,

$$P(\omega)=\mathrm{Re}\left\{\frac{\Delta(\omega)}{[\omega^2-\Delta^2(\omega)]^{1/2}}\right\}$$

Multiplying (5') by $\mathbf{P}^{-1}$, we obtain the linear integral equation

$$g=\mathbf{P}^{-1}\mathbf{B}+\mathbf{P}^{-1}\mathbf{C}\{g\}, \quad (6)$$

which is resolved by the iteration method. Corrections of the experimental function $N_{exp}(\omega)$ are carried out by using the conditions

$$\int_{\Delta_0}^{\infty}d\omega[N(\omega)-N_{BCS}(\omega)]=0,\quad \int_{\Delta_0}^{\infty}N(\nu)\frac{d\nu}{\omega^2-\nu^2}\Bigg|_{\omega\to\Delta_0}\to 0. \quad (7)$$

The Eliashberg $\alpha^2F(\omega)$ function must be positive and go to zero at $\omega>\omega_0$. These (and any another) physical constraints for $\alpha^2F(\omega)$ and $N(\omega)$ are obeyed by the iteration procedure in which the initial experimental density of states $N_{exp}(\omega)=dI/dV_S/dI/dV_N$ is corrected by the simple function $\delta n(\omega,C)=c_1+c_2\omega+c_3\omega+\ldots$. The parameters $C=c_1,c_2,c_3\ldots$ are chosen so that the functional

$$\Psi(C) = \int_{\omega_a}^{\omega_b} \left(n_c(\omega) - \mathbf{EL}\{n_c\}\right)^2 d\omega$$

is minimized. Here

$$n(\omega) = \frac{N(\omega)}{N_{BCS}(\omega)} - 1, \quad n_c(\omega) = n_{exp}(\omega) + \delta n(\omega, C),$$

operator **E** correspond to the equation (1, 2)

$$n(\omega) = \mathbf{E}\{\alpha^2 F\}, \text{ i.e. } \alpha^2 F(\omega) \to \Delta(\omega) \to N(\omega) \to n(\omega)$$

and operator **L** correspond to relation (3), (4) and linear equation (6),

$$\alpha^2 F(\omega) = \mathbf{L}\{n\}. \text{ i.e. } N(\omega) \to n(\omega) \to \Delta(\omega) \to \alpha^2 F(\omega).$$

```
      INTEGER*2 HB,MB,SB,HSB
      DIMENSION SAM(70)
      DIMENSION YG(501),E(501),CNexp(501),CN(501),CF(501),DR(501),
     *DI(501),ZI(501),FI(501),PN(501),PF(501),DP(1151),DM(1151),
     *GP(302),GM(302),C1(151),C2(151),AM(60,60)
      EQUIVALENCE (AM,E)
      COMPLEX UIV

      OPEN(5,FILE='INP.DAN',STATUS='OLD')
      OPEN(6,FILE='OUT.DAN',STATUS='OLD')
      OPEN(1,FILE='A2F.PRN',STATUS='OLD')
      OPEN(2,FILE='Nexp.PRN',STATUS='OLD')

      OPEN(3,FILE='Ncal.PRN',STATUS='OLD')
      OPEN(7,FILE='DR.PRN',STATUS='OLD')
      OPEN(8,FILE='DI.PRN',STATUS='OLD')

      OPEN(9,FILE='ZR.PRN',STATUS='OLD')
      OPEN(10,FILE='ZI.PRN',STATUS='OLD')
      READ(5,5) SAM
 5    FORMAT(70A4)
      WRITE(6,5) SAM
      READ(5,410) JA,JX,J0,JD,JQ,MG
410   FORMAT(6I10)
C     JA - The index of first "judiciously" chosen data point
C     JX - (JX-1)*H=Wo is the maximum phonon energy
C     J0 - Experimental data Nexp(E) are used
```

```fortran
C              up to Emax=(J0-1)*H, J0>JX+2
C      JD - The number of experimental Nexp data points
C              including the zero point
C      JQ - The cutt-off energy Wc=H*(JQ-1)
C      MG - The Eliashberg function a2F(w) is proportional
C              to w^2 up to Emin=H*(MG+1)
       JC=JQ

       WRITE(6,413) JA,JX,J0,JD,JQ,MG
413    FORMAT(' JA=',I4,'  JX=',I4,'  J0=',I4,'  JD=',I4,
      *'  JQ=',I4,'  MG=',I4,/)

       READ(5,411) NIT,MX,H,D,WH,TC,ALAMBD
411    FORMAT(2I10,5F10.3)
C
C      NIT - The number of iterations necessary
C             to fit the computed and experimental N(w) data.
C             If Nit=0 the a2F(w) is determined
C              by the dispersion method directly
C      MX  - The number of iterations to solve the Eliashberg equation
C      H   - The energy spacing in meV for the calculations
C      D   - The energy gap for the superconductor under investigation
C      WH  - The starting value for the Coulomb pseudopotential Mu*
C      TC  - The given value of the critical temperature Tc
C      ALAMBD - The given value of the Eliashberg EPI constant lambda.
C
       WRITE(6,412) NIT,MX,H,D,WH,TC,ALAMBD
412    FORMAT(' NIT=',I4,' MX=',I4,' H=',F8.3,
      *' D=',F8.3,/,' Start Mu=',F8.4,
      *' TC=',F8.3,' Start Lambda =',F8.3,/)

       READ(5,410) INP,NBreg,NBmu,NBinput,Nbzero,NBa2F
C      INP   - If INP=0 then Z(w) and D(w) functions
C               are not printed (they are printed when INP=1).
C      NBreg - The parameter NBreg determines the correction
C               function dN(w) to the experimental Nexp(w):
C         - NBreg=1          N1(w)=C1
C         - NBreg=10         N1(w)=C2+C2*w
C         - NBreg=25-40      N1(w)=<C1+C2*w>
C              Here <> denotes an average values.
C              The program chices the constant C1, C2 automatically.
C      NBmu  - If MBmu=1 the computation are fulfilled
C               with the fixed value of the pseudopotential Mu*C

C      NBinput=1 - input data are experimental tunnel resistance in
C                  normal (Rn) and superconductive (Rs) state
C      NBinp  =2 - input data is the normalized tunnel conductivity
C      NBzero =1 - A setting to the zero values of the
C                  a2F(J) finction at JX<J<J0
C      Nba2F = 1 - Computation with the user-defined function a2F(w)
       IF(NBreg.GT.2) NBzero=1
       IF(NIT.EQ.0) NIT=1
       IF(NBa2F.EQ.1) THEN
        IF(MX.LT.3) MX=3
        NIT=1
       END IF

       DO I=1,501
          CN(I)=0.0
       END DO
       WRITE(6,711) INP,NBreg,NBmu,NBinput,NBzero,NBa2F
```

```fortran
71 1  FORMAT('  INP=',I4,'  NBreg=',I4,'  NBmu=',I4,/,
     *'  NBinput=',I4,'  NBzero=',I4,'  NBa2F=',I4,/)
      NB3=NBinput
C
C     If NB3=1 then use Rn and Rs data;
C     for NB3=2 the input data is the normalized conductivity Nexp
C      for the tunnel S-I-N or S1-I-S2 tunnel junction
C
      IF (NB3.GT.0) THEN
         CALL ONE(D,H,JD,JA,NB3,CN,CF,PN,PF)
      ELSE
        READ(5,14) (CN(I),I=1,JD)
      END IF ! NB3>0

      WRITE(2,7200) (CN(I),I=1,JD)
7200  FORMAT(F10.5)
 14   FORMAT(5F10.5)

      WRITE(6,415) (CN(I),I=1,JD)
415   FORMAT(' INPUT REDUCED DENSITY OF STATES N(w)'/(5F10.5))

      DO I=1,501
       CNexp(I) = CN(I)
      END DO
      JDstart=JD
      NW=0
      NN=0
      IWH=1
      JR=2
      NF1=7
      NGIT=10
      MXA=MX
      LAMBD=0

      IF(NBmu.EQ.1) THEN
       IF(MX.LT.10) MX=10
      END IF

      NU=JC
      IF(JD.EQ.J0) J0=J0-1
      JB=J0+1
      IF(JX.GT.J0) JX=J0
      AI=JQ-1
      XC=D+AI*H
      D2=D*D

      DO I=1,JC
       AI=I-1
       E(I)=D+H*AI
      END DO

      KD=2*JX
      IF(JD.GT.JC) JD=JC
      JC=JD
      A=JA-1
      CN(1)=0.8*CN(JA)
      DO I=1,JA
       AI=JA-I
       BI=I-1
       CN(I)=(CN(1)*AI+CN(JA)*BI)/A
      END DO
```

```
      DO I=1,JB
       CNexp(I)=CN(I)
      END DO

      IF(JC.EQ.NU) GO TO 430
      A0=A
      IF(JD.LT.KD) JD=KD
      IF(JC.GT.KD) GO TO 472
      AJ=JC
      B=CN(JC)*AJ**6
      AJ=AJ-1.0
      A=B-CN(JC-1)*AJ**6
      DO I=JC,KD
       AI=I
       AI=AI**6
       AJ=I-JC
       CN(I)=(A*AJ+B)/AI
      END DO

      JC=KD
472   A=-0.44*D2
      Q=E(JC)
      B=(CN(JC)*Q**2-A)*Q**2

      DO I=JC,NU
       X=E(I)
       X=X*X
       CN(I)=(A*X+B)/(X*X)
      END DO
      JC=NU
C
C     Calculation of the weight coefficients for intergral sum
C
430   Q=E(JC)
      SQ=SQRT(Q*Q-D2)
      IW=JD-1
      JJ=JC-1
      CC=0.0
      ZA=0.0
      II=0
      ZB=0.5*D2*ALOG(D)
      SQL=1.0E-8

      DO 45 I=1,JJ
       X=E(I)
       S=X+H
       SQH=SQRT(S*S-D2)
       C=H*X/SQL

       IF(II.GT.0) GO TO 41

        ELN=ALOG(S+SQH)
        AA=SQH
        BB=0.5*(S*SQH+D2*ELN)
        A=AA-ZA
        B=BB-ZB-(X+0.5*H)*A
        A=A*0.5
        B=B/H
        Z=CC+A-B
        PN(I)=Z
```

```fortran
          CC=A+B
          ZA=AA
          ZB=BB
          A=ABS(Z/C-1.0)
          IF(A.LT.0.0001) II=1
          GO TO 42
41        PN(I)=C
42        SQL=SQH
45     CONTINUE

       PN(JC)=PN(JJ)*0.5
       SQL=1.0E-8
       II=0
       ZB=0.0
       CC=0.0
       ZA=ALOG(D)

       DO 58 I=1,JJ
        X=E(I)
        S=X+H
        SQH=SQRT(S*S-D2)
        C=H/SQL

        IF(II.GT.0) GO TO 54
         AA=ALOG(S+SQH)
         BB=SQH
         A=AA-ZA
         B=(BB-ZB-(X+0.5*H)*A)/H
         A=A*0.5
         Z=A-B+CC
         CC=A+B
         ZA=AA
         ZB=BB
         PF(I)=Z
         A=ABS(Z/C-1.0)
         IF(A.LT.0.0001) II=1

         GO TO 53
54       PF(I)=C
53       SQL=SQH
58     CONTINUE

       PF(JC)=0.5*PF(JJ)
       IMAX=2*JC+JB
       DH=2.0*D/H
       DP(1)=0.5/DH
C
C      Finding of the weight factors for dispersion relations
C
       DO I=2,IMAX
        AI=I-1
        DP(I)=1.0/(AI+DH)
       END DO

       IMIN=JC+1
       DO I=IMIN,IMAX
        AI=I-JC
        DM(I)=1.0/AI
       END DO

       DM(JC)=0.0
```

```
      IMAX=JC-1

      DO I=1,IMAX
       IP=JC+I
       IM=JC-I
       DM(IM)=-DM(IP)
      END DO

 4100 J9=J0+10
C
C     Start for a2F(w) calculation
C
      T9=0.0
      A=J9

      DO I=1,JC
       B=1.0
       AI=I
       IF(I.GT.J9) B=(A/AI)**4
       T9=T9+B*PN(I)
      END DO

      A=CN(JC)*Q*Q/SQ

      DO I=1,JC
       A=A+CN(I)*PN(I)
      END DO

  445 A=-A/T9
      B=J9
C
C     Correction of the density of states N(w) by the sum rules
C
      DO I=1,J9
       CN(I)=CN(I)+A
      END DO
      DO I=J9,JC
       AI=I
       CN(I)=CN(I)+A*(B/AI)**4
      END DO

      IMAX=2
      IF(JR.LT.0) IMAX=IW
      I=JC+1
      DM(I)=1.0
      I=JC-1
      DM(I)=-1.0
      CST=Q*Q*((1.0+CN(JC))*Q/SQ-1.0)
C
C     Calculation of the dispersion integral
C
      DO I=2,IMAX
       IM=JC-I
       S=E(I)
       IP=I-1
       A=CN(I)
       AA=SQRT(S*S-D2)
       B=0.0
        DO J=1,JC
         JP=J+IP
         JM=J+IM
```

```
             B=B+PN(J)*(CN(J)-A)*(DM(JM)-DP(JP))
            END DO
           B=B+PN(I)*(CN(I+1)-CN(I-1))*0.5
           B=0.5*B/(S*H)
           ZA=ALOG((Q-S)/(Q+S))*0.5/S
           ZB=ALOG((SQ-AA)/(SQ+AA))*0.5/AA
           BB=ZB-ZA
           BB=A*ZB-CST/(S*S)*(ZA+1.0/Q)+BB
           B=B+BB
           CF(I)=-B*2.0*AA/3.1415927
          END DO
          IF(JR) 82,81,81
   81     JR=JR-1
          BB=0.0
          K=JA+1
          A=H*H
          ZR1=2.0*D/H+1.0
          CB=0.5*(CN(3)-CN(1))*PN(2)/(1.0+ZR1)
          CB=CB-(CN(1)-CN(2))*PN(1)/ZR1
          DO I=3,JA
           AI=I-1
           CB=CB+PN(I)*(CN(I)-CN(2))/((AI-1.0)*(AI+ZR1))
          END DO

          DO I=K,JC
           AI=I-1
           BB=BB+PN(I)/((AI-1.0)*(AI+ZR1))
          END DO
          CB=CB/A
          BB=BB/A-ZB
          CC=2.0*AA*(CB+(CN(JA)-CN(2))*BB)
          A=1.0+CF(2)/CC*3.1415927
          B=CN(JA)*(1.0-A)
C
C      Correction of the first N(w) values by the condition Im{S(0)}=0
C
          DO I=1,JA
           CN(I)=A*CN(I)+B
          END DO
           GO TO 4100
   82     IF(NN) 86,86,103

   86     AI=IW
          AA=CF(IW)*AI**5
          DO I=IW,JC
           AI=I
           CF(I)=AA/AI**5
          END DO
          AI=JA-1
          AA=CF(JA)*AI**(-2.5)
          DO I=1,JA
           AI=I-1
           CF(I)=AA*AI**2.5
          END DO

          IMAX=JC
  103     Z=1.0
          ZI1=1.0
          ZR1=1.0
C
C      Calculation of the complex energy gap function D(w),
```

```
C          DR=Real{D}, DI=Imagine{D}
C
       DO 90 I=2,IMAX
        S=E(I)
        S2=S*S
        A=CN(I)
        B=CF(I)
        AA=S2*(A*(A+2.0)-B*B)+D2
        A=A+1.0
        BB=2.0*S2*B*A
        UIV=CSQRT(CMPLX(AA,BB))
        ZA=REAL(UIV)
        ZB=AIMAG(UIV)
        C=Z*B
        IF(C) 94,94,95
94      IF(AA) 96,96,97
97      ZI1=-ZI1
        GO TO 95
96      ZR1=-ZR1
95      CC=A*A+B*B
        DRI=ZR1*ABS(ZA*A+ZB*B)/CC
        DII=ZI1*ABS(ZB*A-ZA*B)/CC
        Z=B
        DR(I)=DRI
        DI(I)=DII
90     CF(I)=DRI*A-DII*B

       CF(1)=2.0*CF(2)-CF(3)
       DR(1)=D
       DI(1)=0.0

       ATC=Q*CF(JC)/SQ
       DO I=1,JC
         CN(I)=(CN(I)+1.0)*PN(I)
       END DO
       DO I=1,JC
         CF(I)=CF(I)*PF(I)
       END DO
C
C      Here DR=Re{D) and DI=Im{D}
C
       I=JC+1
       DM(I)=1.5
       I=JC-1
       DM(I)=-1.5
       KMAX=2*JB

       DO K=1,KMAX
        S=0.0
        DO J=1,JC
         IP=K+J
         S=S+CN(J)*DP(IP)
        END DO
        Z=E(K+1)
        B=Z/Q
        A=ALOG(1.0+B)
        AC=A+CST/(Z*Z)*(A-B)
        GP(K)=S-H*AC
       END DO

       C=-0.5*CST/(Q*Q)
```

```
        JJ=JC-JB-1
        DO K=1,KMAX
         S=0.0
         JM=JJ+K
          DO J=1,JC
            IM=JM+J
            S=S+CN(J)*DM(IM)
          END DO
         AI=K-JB
         Z=H*AI-D
         B=Z/Q
         A=ALOG(1.0+B)
         AC=C
         AA=ABS(B)
         IF(AA.GT.0.0001) AC=A+CST/(Z*Z)*(A-B)
         GM(K)=S-H*AC
        END DO

        IF(NN) 130,130,4108
130     S=1.0/CF(1)
        C2(1)=S
        II=JB-1
        DO J=2,II
         C=0.0
         IM=J+1
          DO K=2,J
            JM=IM-K
            C=C+C2(JM)*CF(K)
          END DO
         C2(J)=-C*S
        END DO
        DO K=1,J0
         AJ=((K-1)*(J0-K))**2
         YG(K)=AJ
        END DO
        S=0.0
        DO K=1,J0
         S=S+YG(K)
        END DO
        DO K=1,J0
         YG(K)=YG(K)/S
        END DO
        DO J=2,J0
         S=0.0
         JM=JB-J
         JP=J-2
          DO K=2,J0
            IP=K+JP
            IM=K+JM
            S=S+YG(K)*(GP(IP)-GM(IM))
          END DO
         C=0.0
         JM=J+1
          DO K=2,J
            IM=JM-K
            C=C+YG(K)*CN(IM)
          END DO
         X=E(J)
         C1(J)=(DR(J)*C-DI(J)*S/3.1415927)/X
        END DO
        A=0.0
```

```fortran
              B=0.0
              DO J=2,J0
                JM=J+1
                  DO I=2,J
                    IM=JM-I
                    S=C2(IM)
                    A=A+S*C1(I)
                    B=B+S*DI(I)
                  END DO
              END DO
              S=B/(1.0-A)/3.1415927
              DO K=2,J0
                YG(K)=YG(K)*S
              END DO
C
C     Here YG is the zero approximation for a2F(w)
C
4108          SQL=0.0
              II=1
              JW=JB
              JJ=1
              YG(1)=0.0
              YG(JW)=0.0
              IF(NIT) 151,151,150
151           J0=JX
              JW=JX+1
150           KMIN=MG+1
C
C     Solution of the linear integral equation for a2F(w)
C
4103    DO J=2,JW
                JM=J+1
                C=0.0
                  DO K=2,J
                    IM=JM-K
                    C=C+YG(K)*CN(IM)
                  END DO
                JP=J-2
                JM=JB-J
                S=0.0
                  DO K=2,J0
                    IP=K+JP
                    IM=K+JM
                    S=S+YG(K)*(GP(IP)-GM(IM))
                  END DO
                X=E(J)
                A=DR(J)*C/X+DI(J)*(1.0-S/X)/3.1415927
417             C2(J)=A
              END DO
473     B=CF(1)
              A=C2(KMIN)
              DO J=2,KMIN
                JM=J+1
                C=0.0
                  DO K=2,J
                    AI=(K-1)**2
                    C=C+AI*CF(JM-K)
                  END DO
                C2(J)=C
              END DO
```

```fortran
      A=A/C2(KMIN)
      DO J=2,KMIN
       C2(J)=C2(J)*A
      END DO
      C1(2)=C2(2)/B
      DO J=3,J0
       A=0.0
       L=J-1
       JM=J+1
        DO K=2,L
         IM=JM-K
         A=A+C1(K)*CF(IM)
        END DO
       C1(J)=(C2(J)-A)/CF(1)
      END DO
      DO I=2,J0
       C2(I)=C1(I)-YG(I)
      END DO
      DEL=0.0
      A=J0-MG
      DO I=MG,J0
       DEL=DEL+C2(I)**2
      END DO
      DEL=SQRT(DEL/A)
      C2(1)=0.0
      C2(JW)=0.0
      BNG=0.25
      IF(II.GT.NF1) JJ=0
      IF(JJ) 173,173,172
172   DO K=2,J0
       YG(K)=C1(K)
      END DO
      GO TO 174
173   DO K=2,J0
       YG(K)=YG(K)+(BNG*(C2(K-1)+C2(K+1))+0.5*C2(K))
      END DO
174   A=YG(KMIN+1)
      II=II+1
      AJ=KMIN
      S=2.0

      DO I=1,KMIN
       AI=I-1
       YG(I)=A*(AI/AJ)**S
      END DO
      J=II-NGIT
      IF(J.LT.0) GO TO 4103
      IF(NW) 181,181,182
C
C     Determination of the Coulomb pseudopotential W=Mu*
C     without solution of the Eliashberg equations
C
181   ZA=0.0

      DO J=2,J0
       IP=J-1
       IM=IP+JB
       ZA=ZA+YG(J)*(GP(IP)-GM(IM))
      END DO
      ZA=D-ZA
      KMAX=J0-1
```

```
      DO K=1,KMAX
       S=0.0
        DO J=1,JC
         IP=K+J
         S=S+CF(J)*DP(IP)
        END DO
       Z=E(K+1)
       B=Z/Q
       AC=-ATC/Z*ALOG(1.0+B)
       GP(K)=S-H*AC
      END DO
      KMAX=J0+JB-1
      PSQ=-ATC/Q
      JJ=JC-JB-1
      DO K=JB,KMAX
       S=0.0
       JM=JJ+K
        DO J=1,JC
         IM=JM+J
         S=S+CF(J)*DM(IM)
        END DO
       AI=K-JB
       Z=H*AI-D
       B=Z/Q
       C=ABS(B)
       AC=PSQ
       IF(C.GT.1.0E-4) AC=-ATC/Z*ALOG(1.0+B)
       GM(K)=S-H*AC
      END DO
      C=0.0
      DO J=2,J0
       IP=J-1
       IM=IP+JB
       C=C+YG(J)*(GP(IP)+GM(IM))
      END DO

      PU=ATC*ALOG(XC/Q)
      DO I=1,JC
       PU=PU+CF(I)
      END DO

      W=(C-ZA)/PU
C
C     Here W is the 'dispersion' Coulomb pseudopotential
C
182   A2=0.0
      CC=0.0
      ELOG=0.0
      DO I=2,JX
       AI=I-1
       A=YG(I)
       CC=CC+A/AI
       A2=A2+A
       ELOG=ELOG+(A/AI)*LOG(H*AI)
      END DO
      CC=2.0*CC
      ELOG=EXP(2*ELOG/CC)
      A2=A2*H
      WRITE(6,29) DEL
29    FORMAT('DEL=',E13.6,/)
      WRITE(6,705) A2, ELOG, CC
```

```fortran
705   FORMAT(' A2=',F8.3,' OmegaLog=',F8.3,'  Lambda=',F8.3,/)

      WRITE(6,7208) W
7208  FORMAT(' Coulomb potential (dispersion) Mu=',F8.3)

      IF(Nbzero) 202,202,203
       WRITE(6,28)
28     FORMAT(' The a2F function is corrected',/)
C
C     Some correction of the calculated a2F(w)
C
203   L=1
      A=0.0
      DO I=JX,J0
       A=A+YG(I)
      END DO
      IP=(J0-JX+1)*(J0-L)
      AI=IP
      A=A/IP
      DO I=L,J0
       AI=I-L
       YG(I)=YG(I)-A*AI
      END DO
      KMIN=MG+1
      S=2.0
      A=YG(KMIN+1)
      AJ=KMIN
      DO I=1,KMIN
       AI=I-1
       YG(I)=A*(AI/AJ)**S
      END DO

202   DO I=1,JB
       C1(I)=CN(I)/PN(I)-1.0
      END DO
C
C     Correction of the a2F(w) values (constraints)
C
      DO I=1,J0
       A=YG(I)
       IF(A.LT.0.0) A=0.0
       YG(I)=A
      END DO
      DO I=JX,JB
       YG(I)=0.0
      END DO

      IF(NBa2F.EQ.1) THEN
       LAMBD=1
C
C      Input of the given a2F(w) function for control calculations
C       and for improvement of the regularization process
C
       READ(5,14)(YG(I),I=1,J0)
       A=0.0
       DO I=2,J0
          A=A+YG(I)/(I-1)
       END DO
       A=2.0*A
C
C      ALAMBD is the given lambda value
```

```fortran
C
      DO I=2,J0
         YG(I)=ALAMBD*YG(I)/A
      END DO

      WRITE(6,9205)(YG(I),I=1,J0)
9205  FORMAT(' Initial function a2F(w)='/(5F10.5))
      END IF ! NBa2F=1

4333  CONTINUE
C
C     Solution of the Eliashberg equation
C       for the given function YG=a2F(w)
C
C     START
C
      PU=ATC*ALOG(XC/Q)

      DO J=1,JC
         PU=PU+CF(J)
      END DO

      DO I=2,JC
       ZA=0.0
       ZB=0.0
       KMAX=I
       IF(I.GT.JX) KMAX=JX
       JM=I+1

         DO K=1,KMAX
            A=YG(K)
            IM=JM-K
            ZA=ZA+CN(IM)*A
            ZB=ZB+CF(IM)*A
         END DO

       X=E(I)
       ZI(I)=ZA/X
       FI(I)=ZB
      END DO

      ZI(1)=0.0
      FI(1)=0.0
      K=JC-1
      A=ZI(JC)*Q
      B=FI(JC)*Q

      DO I=1,K
       X=E(I)
       AA=ALOG((Q+X)/(Q-X))/X
       ZB=B*AA
       ZA=A*AA
       IP=I-1
       IM=JC-I

          DO J=2,JC
            JP=IP+J
            JM=IM+J
            AA=DP(JP)+DM(JM)
            ZA=ZA+ZI(J)*AA
            ZB=ZB+FI(J)*AA
```

```fortran
           END DO

        CN(I)=ZA
        CF(I)=ZB
       END DO

       B=CF(1)-D*(1+CN(1))
       W=B
       A=1.0
       CN(JC)=CN(K)
       CF(JC)=CF(K)

       IF(LAMBD.GT.0.AND.NBmu.EQ.0) GO TO 230
       IF(IWH) 230,230,231

231    B=WH*PU
       A=(D+B)/(CF(1)-D*CN(1))

       DO I=1,JX
        YG(I)=YG(I)*A
       END DO

230    AI=A*3.1415927
       DO I=1,JC
         CN(I)=1.0+A*CN(I)
         ZI(I)=ZI(I)*AI
         CF(I)=A*CF(I)-B
         FI(I)=FI(I)*AI
       END DO

       MAK=0
       IF(IWH.GT.0) MAK=1
       IF(MX.GT.0) MAK=1
       IF(LAMBD.GT.0) MAK=1
       DC=0.0
       DA=0.0

       IF(MX.GT.0) GO TO 238

       IF((NIT-1).EQ.0.AND.INP.EQ.1) THEN
        WRITE(9,7200) (CN(I),I=1,JC)
        WRITE(10,7200) (ZI(I),I=1,JC)
C
C      Here CN and ZI are the real and imagine parts
C        of the renormalization function Z(w)
C
       END IF
238    CONTINUE
C
C      Calculation of Re{D}, Im{D} and density of states N(w)
C
       DO 5233 I=2,JC
        ZA=CN(I)
        AA=CF(I)
        BB=FI(I)
        ZB=ZI(I)
        A=ZA*ZA+ZB*ZB
        Z2=A
        DRI=(ZA*AA+ZB*BB)/A
        DII=(ZA*BB-ZB*AA)/A
        X=E(I)
```

```
            X2=X*X
            AA=ABS(X2-DRI*DRI+DII*DII)
            BB=-DRI*DII
            BB=BB+BB
            UIV=CSQRT(CMPLX(AA,BB))
            U=REAL(UIV)
            V=AIMAG(UIV)
            IF(U) 240,241,241
240         U=-U
            V=-V
241         A=SQRT(X2-D2)/(U*U+V*V)
            CNI=A*U-1.0
            IF(MAK.GT.0) GO TO 2003

            IF(I.GT.JB) GO TO 1000
            IF(I.LT.JA) GO TO 5233
            DDR=((DRI-D)*ZA+DII*ZB)/Z2
            DDI=(DII*ZA-(DDR-D)*ZB)/Z2
            CII=(DRI*DDR-DII*DDI)/(X2-D2)
            IF(I.EQ.JA) GO TO 8241
            B=CII-CIJ
            DC=DC+B**2
            DA=DA+(C1(I)-C1(I-1)-CNI+CNJ)*B
8241        CIJ=CII
            CNJ=CNI
            GO TO 5233
2003        CN(I)=CNI
            CF(I)=A*(U*DRI+V*DII)
            DR(I)=DRI
            DI(I)=DII
5233        CONTINUE

            IF(MAK.GT.0) GO TO 4000
1000        MAK=MAK+1
            A=1.0+DA/DC
            DO I=1,JX
             YG(I)=YG(I)*A
            END DO
            B=D*(1.0-A)
            W=A*W-B

            DO I=1,JC
             ZI(I)=ZI(I)*A
             FI(I)=FI(I)*A
             CN(I)=1.0+A*(CN(I)-1.0)
             CF(I)=B+A*CF(I)
            END DO

            DC=0.0
            DA=0.0
            GO TO 238
4000        DR(1)=D
            CF(1)=2.0*CF(2)-CF(3)
            CN(1)=2.0*CN(2)-CN(3)
            PU=Q*CF(JC)/SQ*ALOG(XC/Q)

            DO J=1,JC
             PU=PU+CF(J)*PF(J)
            END DO
            W=W/PU
C
```

```fortran
C     W- The Eliashberg Coulomb potential Mu*
C
      DR(1)=D
      DI(1)=0.0

      IF(MX.GT.0) THEN
       MX=MX-1
       ATC=Q*CF(JC)/SQ
       DO I=1,JC
        CN(I)=(CN(I)+1.0)*PN(I)
        CF(I)=CF(I)*PF(I)
       END DO

       GO TO 4333
      END IF
c
C     FINISH
C
C     The end of the Eliashberg equations solution
      DO I=1,JB
       C2(I)=CNexp(I)-CN(I)
      END DO

      B=0.0
      DO I=JA,JB
       B=B+(C1(I)-CN(I))**2
      END DO
      AI=JB-JA
      EIT=SQRT(B/AI)
      A=0.0
      DO I=JA,JB
       A=A+CN(I)-CNexp(I)
      END DO

      AIT=A/AI

      WRITE(6,8251) EIT
8251  FORMAT(' Square root error for N(W)',E13.3,/)
      WRITE(6,7208) W

      WRITE (6,8252) AIT
8252  FORMAT(' A difference between the experimental-mesaured'
     *,/,' and computed density of states.   AIT=',E13.3,/)

      NF1=4
      NGIT=6
      IW=JB
      JR=2
      NBa2F0=NBa2F

      IF(NBa2F.EQ.1) THEN
       NBa2F=0
       GO TO 296
      END IF
C
C     Procedure REG produce the correction
C      of the experimental function N(w)exp
C
      IF(NIT.GT.0) CALL REG(NBreg,JA,JB,CN,CNexp,C2,D)

296   IWH=0
```

```fortran
      IF(NBmu.EQ.1.AND.NIT.EQ.2) THEN
       IWH=1
       MX=10
      END IF
      NN=1
      NIT=NIT-1
      NW=NIT

      IF(NIT.GT.0) GO TO 4100

 4101 CONTINUE
C
C     Calculated reduced density of states
C
      WRITE(3,7200) (CN(I),I=1,JDstart)

      IF(INP.EQ.1) THEN
       WRITE(7,7200) (DR(I),I=1,JC)
       WRITE(8,7200) (DI(I),I=1,JC)
      END IF
C
C     Calculated a2F(w)
C
      WRITE(1,7200) (YG(I),I=1,J0)

      A2=0.0
      CC=0.0
      ELOG=0.0
      DO I=2,JX
       AI=I-1
       A=YG(I)
       CC=CC+A/AI
       A2=A2+A
       ELOG=ELOG+(A/AI)*LOG(H*AI)
      END DO
      CC=2.0*CC
      ELOG=EXP(2*ELOG/CC)
      A2=A2*H

      WRITE(6,7) W
 7    FORMAT(' The Eliashberg Coulomb potential MU=',F10.3,/)

      WRITE(6,33)
 33   FORMAT('The Eliashberg value are:',/)

      WRITE(6,705) A2,ELOG,CC

      IF(TC.GT.0.001) THEN
       AJ=(XC/0.27/TC+1.0)*0.5
       AI=AJ
       IF(AJ.GT.59.0) AJ=59.0
       M=AJ
       W=W/(1.0+W*ALOG(AI/AJ))
C
C     Calculation of the critical temperature Tc
C
       CALL FTC(H,W,J0,M,TC,AM,YG)
       WRITE(6,315) TC
 315   FORMAT(' Critical temperature Tc=',F8.3)
      END IF ! Tc>0
```

```fortran
      IF(NBa2F0.EQ.1) THEN
       WRITE(6,25)
 25    FORMAT(' a2(w)F(w) function was given',/)
       IF(NBmu.EQ.0) then
        WRITE(6,27)
 27     FORMAT(' The lambda value was fixed',/)
       END IF
      END IF

      IF(NBmu.EQ.1) THEN
       WRITE(6,26)
 26    FORMAT(' The Mu value was fixed',/)
      END IF

      STOP
      END

      SUBROUTINE ONE (Delta0,Hcalc,JC,JA,NB3,CN,RN,RS,SIGMA)

      DIMENSION  CN(1),RN(1),RS(1),SIGMA(1)
C     Calculation of the reduced tunnel density of states
C
C     Delta0 - the value of energy gap of the given superconductor
C     Hcalc  - the energy spacing in meV for the calculations
C     JC -     the number of calculated data points
C     JA -     the index of first "judiciously" chosen data point
C     NB3 =1 - input data are experimental tunnel resistance in
C              normal (Rn) and superconductive (Rs) state
C     NB3 =2 - input data is the normalized tunnel conductivity
C     CN -     the experimental normalized density of states
C
      READ(5,10) Nexp, Hexp, DeltaR, Sigma0, Ustart, DeltaInj
 10      FORMAT((6X,I4),5F10.5)

C     Nexp -    the number of experimental points in Rn and Rs
C     Hexp -    the spacing between data points in mV
C     DeltaR -  shift of the Rn curve relatively
C               to Rs- curve at the last data point
C     Sigma) -  the tunnel conductance at zero point
C     Ustart -  the initial data bias from which
C               the experimental data are used
C     DeltaInj -the superconductor injector energy gap
C
      IF(NB3.EQ.1) THEN
        READ(5,20) (RN(I),I=1,Nexp)
 20     FORMAT(5F10.5)
        READ(5,20) (RS(I),I=1,Nexp)
        WRITE(6,60) (RN(I),I=1,Nexp)
 60     FORMAT( 'RN=',/(5F10.5) )
        WRITE(6,70) (RS(I),I=1,Nexp)
 70     FORMAT( 'RS=',/(5F10.5) )

        IF (DeltaR.EQ.0.0)  DeltaR=RS(Nexp)-RN(Nexp)
        DO J=1,Nexp
          RN(J) = 1.0/(RN(J)+DeltaR)
          RS(J) = 1.0/RS(J)
          SIGMA(J) = (RS(J)-Sigma0)/(RN(J)-Sigma0)
        END DO
      END IF ! NB3=1
      IF(NB3.EQ.2) THEN
        READ(5,20) (SIGMA(I),I=1,Nexp)
```

```fortran
            WRITE(6,80) (SIGMA(I),I=1,Nexp)
   80       FORMAT( 'NORMALIZED CONDUCTIVITY=',/(5F10.5))
         END IF ! NB3=2

         WRITE(6,40) Nexp, Hexp, DeltaR
   40        FORMAT(' Nexp=',I4,' Hexp=',F6.3,' DeltaR=',F6.3)
         WRITE(6,50) Sigma0, Ustart, DeltaInj
   50        FORMAT(' Sigma0=',F6.3,' Ustart=',F6.3,' DeltaInj=',F6.3)
         DeltaSUM = Delta0 + DeltaInj
         DO J=JA,JC
            Ej = DeltaSUM + Hcalc*(J-1)
            I =INT((Ej-Ustart)/Hexp)
            IF(I.LT.1) I=1
            Ui = Ustart + Hexp*(I-1)
            Ui1 = Ui + Hexp
            CN(J) = (SIGMA(I+1)*(Ej-Ui) + SIGMA(I)*(Ui1-Ej))/Hexp
         END DO
         DO J=1,JA
            CN(J)=CN(JA)
         END DO
         IF(DeltaInj.GT.0.0) THEN
          READ(5,81) CNA
C      CNA is the initial point for the normalized density of states.
C           If you do not know this parameter then set CNA=0.0
   81    FORMAT(F10.3)
          WRITE(6,82) CNA
   82    FORMAT(' CNA=',E13.6)

            CALL TWO(JC,Delta0,DeltaInj,Hcalc,CN,JA,RN,RS,CNA)
         ELSE
         DO J=1,JC
            Ej=Delta0+Hcalc*(J-1)
            CN(J)=CN(J)*SQRT(Ej**2-Delta0**2)/Ej-1.0
         END DO
         END IF

         RETURN
         END
         SUBROUTINE TWO(JC,D1,D2,H,CN,JA,CF,PN,CNA)
C
C      Determination of the tunnel density of states N(w)
C      for S1-I-S2 tunnel junction
C
C      JC - the number of calculated data points
C      D1 - the experimental value of energy gap for S1
C      D2 - the BCS superconductor S2 energy gap
C      H -  the voltage spacing in mV for the calculations
C      CN - the calculated normalized tunnel density of states
C      Ja - the index of first "judiciously" chosen data point
C      CNA -the initial point for the normalized density of states.
C
         DIMENSION CN(1),CF(1),PN(1),BCS(501)
         DIMENSION D(3)

         D(1)=D1
         D(2)=D2
         D(3)=D1
         AH=H
         N=40
         DO L=2,JC
          AI=L-1
```

```
      V=AI*AH+D1+D2
      X=1.0/SQRT(1.0-(D1/(V-D2))**2)+1.0/SQRT(1.0-(D2/(V-D1))**2)-1.0
      V1=(V+D1-D2)/2.0
      V2=(V+D2-D1)/2.0
      X=X+(1.0/SQRT(1.0-(D1/V1)**2)-1.0)*(1.0/SQRT(1.0-(D2/V2)**2)-1.0)
      AN=N
      H=0.5*(V-D1-D2)/AN
      AMP=1.0
      JMAX=2
      IF( (ABS(D1-D2)/(D1+D2)-0.001) .LE. 0.0 ) THEN
        AMP=2.0
        JMAX=1
      END IF
      DO J=1,JMAX
        G1=D(J)
        G2=D(J+1)
        B0=-G1
        DO I=1,N
          AI=I
          EH=G1+AI*H
          EM=EH-0.5*H
          BN=SQRT(EH**2-G1**2)-EH
          X=X-AMP*G2**2*(BN-B0)*((V-EM)**2-G2**2)**(-1.5)
          B0=BN
        END DO
      END DO
     BCS(L)=X
     PN(L)=CN(L)/X-1.0
    END DO
    H=AH
    PN(1)=0.8*CN(JA)
    IF( (CNA-0.0001) .GT. 0.0 ) THEN
     PN(1)=CNA
    END IF
    AJ=JA-1
    DO I=1,JA
     AI=I-1
     BI=JA-I
     PN(I)=(PN(1)*BI+PN(JA)*AI)/AJ
    END DO
    SQ=PN(1)
    CN(1)=0.0
    DO J=2,JC
     CN(J)=(PN(J)-SQ)*BCS(J)
    END DO
    JC=JC-1
    DO J=1,JC
     CN(J)=CN(J+1)
    END DO
    SQ=SQ+1.0
    DO I=2,JC
     S=0.5*CN(I)
     K=I-1
      DO J=1,K
       S=S+CN(J)
      END DO
     CF(I)=S
    END DO
    JC=JC-1
    CF(1)=0.0
    DO I=1,JC
```

```fortran
         CN(I)=CF(I)
        END DO
        B0=0.0
        DO I=1,JC
         AI=I
         X=AI*H
         BN=SQRT(X*(X+2.0*D2))
         CF(I)=(BN-B0)/H
         B0=BN
        END DO
        V=CF(1)
        DO I=1,JC
         S=CN(I)/V
           DO K=I,JC
             J=K-I+1
             CN(K)=CN(K)-S*CF(J)
           END DO
         CF(J)=S
        END DO
        DO I=1,JC
         AI=I
         X=H*AI-0.5*H
         J=JC+1-I
         CN(I)=CF(J)*SQRT(X*(X+2.0*D1))/(X+D1)
        END DO
        JC=JC+1
        DO J=2,JC
         I=JC-J+2
         CN(I)=CN(I-1)
        END DO
        K=JC+1
        CN(K)=2.0*CN(JC)-CN(JC-1)
        SQ=SQ-1.0
        CN(1)=SQ
        DO I=2,JC
         CN(I)=SQ+0.5*(CN(I)+CN(I+1))
        END DO
        JC=JC+1
        CN(JC)=2.0*CN(JC-1)-CN(JC-2)

        RETURN
        END

        SUBROUTINE REG(NBreg,JA,JB,CNcalc,CNexp,deltaCN,D)
        DIMENSION CNcalc(1),CNexp(1),deltaCN(1)
C
C       The subroutine determine the value of parameters
C           C1, C2, C3.. which minimize a difference between
C           the calculated Ncalc(w) and experimental Nexp(w) density
C           of states. This subroutine eliminate the experimental
C           error in the Rn/Rs ratio.
C       If NBreg=2 then Nexp -> Nexp + C1
C
        D2=D*D
        CA=CNcalc(JA)
        CB=CNcalc(JB)
        AJ=JB-JA
        N=Nbreg

        IF(N.EQ.2) THEN
         A=0.0
```

```fortran
        DO J=JA,JB
           A=A+CNcalc(J)-CNexp(J)
        END DO
        DCN=A/AJ
        DO J=JA,JB
           CNcalc(J)=CNexp(J)+DCN
        END DO
       GO TO 3
       END IF

       IF (N.EQ.10) THEN
        A=(CNcalc(JB)-CNexp(JB))/AJ
        B=(CNcalc(JA)-CNexp(JA))/AJ
        DO J=JA,JB
           AA=J-JA
           BB=JB-J
           CNcalc(J)=CNexp(J)+A*AA+B*BB
        END DO
        GO TO 5
       END IF
       IF (N.GT.20) THEN
        NFI=N-20
        L=JA+1
        K=JB-1
         DO J=JA,JB
          deltaCN(J)=CNexp(J)-CNcalc(J)
         END DO
         DO I=1,NFI
          A=deltaCN(JA)
          B=deltaCN(L)
          deltaCN(JA)=(A*2.0+B)/3.0
           DO J=L,K
             C=deltaCN(J+1)
             deltaCN(J)=(A+2.0*B+C)*0.25
             A=B
             B=C
           END DO
          deltaCN(JB)=(A+2.0*B)/3.0
         END DO
         DO J=JA,JB
          CNcalc(J)=CNexp(J)-deltaCN(J)
         END DO
        GO TO 3
       END IF
3      Jmax=1.2*(JB-1)
       L=JB+1
       AI=Jmax-JB
       A=(CB-CNcalc(JB))/AI
       DO J=L,Jmax
          AJ=Jmax-J
          CNcalc(J)=CNcalc(J)-A*AJ
       END DO
       Jmin=JA-1
       AK=Jmin
       A=(CA-CNcalc(JA))/AK
       DO J=1,Jmin
        AJ=J-1
        CNcalc(J)=CNcalc(J)-A*AJ
       END DO
5      CONTINUE
       RETURN
```

```fortran
      END

      SUBROUTINE FTC(H,W,J0,M,TC,AM,YG)
C     Calculation of the critical temperature Tc
C
C     H - the energy spacing in meV
C     W - the Coulomb potential Mu*
C     J0- the index specifying the maximum phonon energy Wo=H*J0
C     M - the matrix dimension
C     TC- the critical temperature
C     YG- Eliashberg electron-phonon function a2F(w)
C
      DIMENSION AM(M,M),YG(J0)
      DIMENSION GG(202),LC(50),MC(50)
      EPS=0.001
C     EPS- RELATIVE ERROR
      TMAX=400
C     TMAX- MAXIUM OF TC
      IT=0
      MAXIT=20
C     MAXIT- MAXIUM OF ITERATION
      TB=1.3*TC
      TA=0.7*TC
      DO I=2,J0
       AI=I-1
       YG(I)=YG(I)*AI
      END DO
  30  T=TA
      K=0
      GO TO 50
  40  T=(TA+TB)*0.5
  50  L=2*M+2
      ZA=(0.54035*T/H)**2
      DO J=1,L
       S=0.0
       AJ=J-1
       A=AJ*AJ*ZA
        DO I=2,J0
          AI=I-1
          S=S+YG(I)/(AI*AI+A)
        END DO
       GG(J)=2.0*S
      END DO
      A=2.0*W
      DO I=1,M
       S=0.0
       IF(I.EQ.1) GO TO 90
       DO J=2,I
        S=S+GG(J)
       END DO
  90   AI=I-1
       S=GG(1)+2.0*(S+AI)+1.0
        DO J=1,I
          AA=GG(I-J+1)+GG(I+J)-A
          AM(J,I)=AA
          AM(I,J)=AA
        END DO
       AM(I,I)=AM(I,I)-S
      END DO
      CALL MINV(AM,M,CT,LC,MC)
      IF(K-1) 120,120,170
```

```fortran
  120 IF(K) 130,130,140
  130 K=1
      AT=CT
      T=TB
      GO TO 50
  140 BT=CT
      AB=SIGN(1.0,AT)*SIGN(1.0,BT)
      IF(AB)160,160,150
  150 A=0.5*ABS(TB-TA)
      TA=TA-A
      IF(TA.LT.0.0) TA=0.0
C     TA MUST BE LARGE THAN 0
      TB=TB+A
      IF(TMAX-TB) 230,230,30
  230 TC=TMAX
      GO TO 210
  160 K=2
      GO TO 40
  170 AB=SIGN(1.0,CT)*SIGN(1.0,AT)
      IF(AB) 180,190,190
  180 TB=T
      BT=CT
      GO TO 200
  190 TA=T
      AT=CT
  200 TC=(TB+TA)/2.0
      IT=IT+1
      IF(IT.GT.MAXIT) GO TO 210
      IF((TB-TA).GT.EPS*TC) GO TO 40
  210 CONTINUE
      DO I=2,J0
       AI=I-1
       YG(I)=YG(I)/AI
      END DO
      RETURN
      END

      SUBROUTINE MINV(A,N,D,L,M)
C       DESCRIPTION OF PARAMETERS
C
C       A - INPUT MMATRIX
C       N - ORDER OF MATRIX A
C       D - RESULTANT DETERMINANT
C       L - WORK VEKTOR OF LENGS N
C       M - WORK VEKTOR OF LENGS N
C
C       REMARKS
C       MATRIX A MUST BE A GENERAL MATRIX
C       METHOD
C       THE STANDART GAUSS-JORDAN METHOD IS USED.
C       THE DETERMINANT OF ZERO INDICATES THAT
C       THE MATRIX IS SINGULAR.
C
      DIMENSION A(1),L(1),M(1)
C
C        SEARCH FOR LARGEST ELEMENT
C
      D=1.0
      NK=-N
      DO 80 K=1,N
       NK=NK+N
```

```
       L(K)=K
       M(K)=K
       KK=NK+K
        BIGA=A(KK)
        DO 20 J=K,N
       IZ=N*(J-1)
        DO 20 I=K,N
        IJ=IZ+I
   10  IF( ABS(BIGA)-ABS(A(IJ))) 15,20,20
   15  BIGA=A(IJ)
       L(K)=I
       M(K)=J
   20  CONTINUE
C
C            INTERCHANGE ROWS
C
        J=L(K)
        IF(J-K) 35,35,25
   25  KI=K-N
       DO 30 I=1,N
        KI=KI+N
        HOLD=-A(KI)
        JI=KI-K+J
        A(KI)=A(JI)
   30  A(JI)=HOLD
C
C            INTERCHANGE COLUMNS
C
   35  I=M(K)
       IF(I-K) 45,45,38
   38  JP=N*(I-1)
       DO 40 J=1,N
       JK=NK+J
       JI=JP+J
       HOLD=-A(JK)
       A(JK)=A(JI)
   40  A(JI)=HOLD
C
C            DIVIDE COLUMN BY MINUS PIVOT (VALUE OF
C            PIVOT ELEMENT IS CONTAINED IN BIGA)
C
   45  IF(BIGA) 48,46,48
   46  D=0.0
       RETURN
   48  DO 55 I=1,N
       IF(I-K) 50,55,50
   50  IK=NK+I
       A(IK)=A(IK)/(-BIGA)
   55  CONTINUE
C
C            REDUCE MATRIX
C
       DO 65 I=1,N
        IK=NK+I
        HOLD=A(IK)
        IJ=I-N
        DO 65 J=1,N
        IJ=IJ+N
        IF(I-K) 60,65,60
   60  IF(J-K) 62,65,62
   62  KJ=IJ-I+K
```

```
         A(IJ)=HOLD*A(KJ)+A(IJ)
   65    CONTINUE
C
C           DIVIDE ROW BY PIVOT
C
         KJ=K-N
         DO 75 J=1,N
            KJ=KJ+N
         IF(J-K) 70,75,70
   70    A(KJ)=A(KJ)/BIGA
   75    CONTINUE
         D=D*SIGN(1.,BIGA)
         A(KK)=1.0/BIGA
   80    CONTINUE
         RETURN
         END
```
Example for INP.DAN :

Lead (Pb) W.L. MacMillan, J.M.Rowell, Phys.Rev.Lett., 14, 516 (1965)
```
         6         109        111        112        501          6
         1          10        0.1        1.4      0.131       7.30          1.0
         0           2          0          0          0          0
    0.0341     0.0341     0.0341     0.0341     0.0341
    0.0341     0.0349     0.0357     0.0364     0.0370
    0.0375     0.0381     0.0387     0.0394     0.0401
    0.0410     0.0420     0.0431     0.0443     0.0454
    0.0464     0.0473     0.0481     0.0489     0.0499
    0.0509     0.0522     0.0537     0.0554     0.0575
    0.0600     0.0629     0.0655     0.0675     0.0684
    0.0684     0.0675     0.0652     0.0616     0.0575
    0.0537     0.0503     0.0470     0.0431     0.0379
    0.0316     0.0260     0.0212     0.0164     0.0119
   0.00766    0.00382   0.000714   -0.00168   -0.00369
  -0.00549   -0.00697   -0.00800   -0.00866   -0.00914
  -0.00950   -0.00977   -0.00982   -0.00965   -0.00940
  -0.00918   -0.00903   -0.00895   -0.00891   -0.00883
  -0.00866   -0.00845   -0.00821   -0.00792   -0.00748
  -0.00684   -0.00608   -0.00536   -0.00484   -0.00462
  -0.00476   -0.00547   -0.00727    -0.0112    -0.0180
   -0.0270    -0.0363    -0.0443    -0.0502    -0.0540
   -0.0558    -0.0560    -0.0553    -0.0542    -0.0530
   -0.0520    -0.0510    -0.0501    -0.0491    -0.0482
   -0.0473    -0.0464    -0.0456    -0.0448    -0.0440
   -0.0433    -0.0426    -0.0419    -0.0413    -0.0407
   -0.0401    -0.0397
```

Example for OUT.DAN:

Lead (Pb)  W.L. MacMillan, J.M. Rowell, Phys. Rev. Lett., 14, 516 (1965)

 JA=   6  JX= 109  J0= 111  JD= 112  JQ= 501  MG=   6

 NIT=   1 MX=  10 H=    .100 D=   1.400
 Start Mu=   .1310 TC=   7.300 Start Lambda =    1.000

  INP=   1  NBreg=    2  NBmu=    0
  NBinput=   0  NBzero=   0  NBa2F=   0

 INPUT REDUCED DENSITY OF STATES N(w)
     .03410     .03410     .03410     .03410     .03410
     .03410     .03490     .03570     .03640     .03700

```
       .03750      .03810      .03870      .03940      .04010
       .04100      .04200      .04310      .04430      .04540
       .04640      .04730      .04810      .04890      .04990
       .05090      .05220      .05370      .05540      .05750
       .06000      .06290      .06550      .06750      .06840
       .06840      .06750      .06520      .06160      .05750
       .05370      .05030      .04700      .04310      .03790
       .03160      .02600      .02120      .01640      .01190
       .00766      .00382      .00071     -.00168     -.00369
      -.00549     -.00697     -.00800     -.00866     -.00914
      -.00950     -.00977     -.00982     -.00965     -.00940
      -.00918     -.00903     -.00895     -.00891     -.00883
      -.00866     -.00845     -.00821     -.00792     -.00748
      -.00684     -.00608     -.00536     -.00484     -.00462
      -.00476     -.00547     -.00727     -.01120     -.01800
      -.02700     -.03630     -.04430     -.05020     -.05400
      -.05580     -.05600     -.05530     -.05420     -.05300
      -.05200     -.05100     -.05010     -.04910     -.04820
      -.04730     -.04640     -.04560     -.04480     -.04400
      -.04330     -.04260     -.04190     -.04130     -.04070
      -.04010     -.03970
 DEL=   .818841E-03

   A2=    3.920 OmegaLog=    4.701  Lambda=    1.535

  Coulomb potential (dispersion) Mu=     .196
  Square root error for N(W)      .693E-02

  Coulomb potential (dispersion) Mu=     .131
  A difference between the experimental-measured
  and computed density of states.   AIT=    -.441E-03

  The Eliashberg Coulomb potential MU=      .131

 The Eliashberg value are:

   A2=    4.041 OmegaLog=    4.702  Lambda=    1.582

   Critical temperature Tc=    7.302
```